%% file: main.tex
\documentclass[twocolumn,reprint,amsmath,amssymb,aps,prl,superscriptaddress,showpacs]{revtex4-1}

\usepackage{multirow}
\usepackage{color}
\usepackage{xcolor}
\usepackage{verbatim}
\usepackage{graphicx}% Include figure files

\usepackage{xspace,amsmath}
\usepackage{dcolumn}% Align table columns on decimal point
\usepackage{bm}% bold math
\newcommand*{\sPlot}{\ensuremath{{}_s{\mathcal P}\mkern-2mu\mathit{lot}}\xspace}

\newcommand{\mev}{\ensuremath{{\mathrm{\,Me\kern -0.1em V}}}\xspace}
\newcommand{\gev}{\ensuremath{{\mathrm{\,Ge\kern -0.1em V}}}\xspace}
\newcommand{\mevcc}{\ensuremath{{\mathrm{\,Me\kern -0.1em V}}/c^2}\xspace}
\newcommand{\gevcc}{\ensuremath{{\mathrm{\,Ge\kern -0.1em V}}/c^2}\xspace}
\newcommand{\gevsq}{\ensuremath{{\mathrm{\,Ge\kern -0.1em V}}^2}\xspace}
\newcommand{\nospacegev}{\ensuremath{{\mathrm{\,Ge\kern -0.1em V}}}}

\newcommand{\gtw}{\texttt{g12}}

\usepackage{soul,color}
\definecolor{chromeyellow}{rgb}{1.0, 0.65, 0.0}
\definecolor{applegreen}{rgb}{0.55, 0.71, 0.0}
\definecolor{asparagus}{rgb}{0.53, 0.66, 0.42}

\begin{document}

%\preprint{APS/123-QED}

\title{First measurement of direct photoproduction of the $a_2(1320)^0$ meson on the proton}% Force line breaks with \\
%\thanks{A footnote to the article title}%

\input{authors.tex}

\date{\today}% It is always \today, today,
             %  but any date may be explicitly specified

\begin{abstract}
We present the first measurement of the reaction $\gamma p \rightarrow a_2(1320)^0 \, p$ in the photon energy range $3.5$--$5.5\gev$ and four-momentum transfer squared $0.2<-t<2.0\gevsq$. Data were collected with the CLAS detector at the Thomas Jefferson National Accelerator Facility. The $a_2$ resonance was detected by measuring the reaction $\gamma p \rightarrow \pi^0 \eta p$ and reconstructing the $\pi^0 \eta$ invariant mass. The most prominent feature of the differential cross section is a dip at $-t\approx0.55\gevsq$. This can be described in the framework of Regge phenomenology, where the exchange degeneracy hypothesis predicts a zero in the reaction amplitude for this value of the four-momentum transfer.
\end{abstract}

\maketitle

It has been more than forty years since Quantum Chromodynamics (QCD) was postulated as the theory of strong interactions. While much progress has been made in understanding the high energy phenomena through this theory, perturbative methods fail to describe the strong interaction at low energies. A clear understanding of this regime is of key importance, since it corresponds to the dominant manifestation of the strong force in nature, in terms of hadrons that constitute the bulk of the visible mass of the Universe.

Hadron spectroscopy is a valuable tool to investigate this regime. The measurement of the meson spectrum, searching for exotic states not compatible with the Quark Model, would provide access to the gluonic degrees of freedom that contribute to the quantum numbers of the hadrons. Investigating the properties and interactions of gluons is critical, since their dynamics give rise to the strong interaction that binds the hadrons.
In this context, the photoproduction of a $\pi^0 \,\eta$ pair on the proton ($\gamma p \rightarrow \pi^0 \eta \,p$) is one of the most promising reaction channels, since any $P$-wave resonance would be unambiguously interpreted as an \textit{exotic}, non $q\overline{q}$ state.
So far, only a few results have been reported for this reaction. At low energies, in the fully non-perturbative regime, high-quality cross-section data have been collected by the GRAAL~\cite{Ajaka:2008zz}, Crystal Ball, TAPS, and A2~\cite{KASHEVAROV2010551,Kashevarov2009}, and CB-ELSA~\cite{Horn:2007pp,Horn:2008qv} collaborations.
In the multi-GeV photon beam energy range, optimal for meson spectroscopy, instead, no data have been published so far.

%Understanding the $\pi^0\,\eta$ production is crucial for spectroscopy, especially for disentangling new resonance signals from non-resonant backgrounds.  
In this energy regime, the $a_2(1320)$ meson is expected to make the dominant contribution to the $\pi^0\,\eta$ invariant-mass spectrum~\cite{Adolph:2014rpp}. It can be thus taken as the reference state for a Partial Wave Analysis of this channel, for example allowing for the interpretation of the variations of the $P-D$ phase difference as a signature for the existence of exotic resonances~\cite{Rodas:2018owy,Albrecht:2019ssa}. Photoproduction of the charged $a_2$ resonance has been measured at SLAC~\cite{PhysRevD.48.3045,PhysRevD.41.3317,Ballam:1969jv}. However, to the best of our knowledge the neutral $a_2$ channel has never been studied in photoproduction.

In this work we report the first measurement of the neutral $a_2(1320)$ meson photoproduction on the proton, for photon beam energies between $3.5$ and $5.5\gev$, and four-momentum transferred squared ($-t$) in the range $0.2\textendash2.0\gevsq$. The differential cross section $d\sigma/dt$ was obtained by measuring the cross section  $d^2\sigma/dt dM$ for the exclusive production of a $\pi^0\, \eta$ pair on the proton, where $M$ is the two-meson invariant mass, and extracting the contribution of the $a_2$ resonance in each kinematic bin. The measurement was performed with the CEBAF Large Acceptance Spectrometer (CLAS) in Hall B at Jefferson Laboratory in a dedicated high-energy, high-statistics run, \gtw.

The experiment used a bremsstrahlung photon beam produced by the interaction of the primary $E_0=5.72$ GeV electron beam with a converter of $10^{-4}$ radiation lengths. A magnetic spectrometer (photon tagger) with energy resolution $0.1 \% E_0$ was used to tag photons in the energy range $0.2E_0$--$0.95 E_0$~\cite{tagger1,tagger2}. The target was a 40-cm-long cell filled with LH$_2$. During the run, the high-intensity photon flux, $\approx 4\times10^7$ $\gamma/$s, was measured by sampling the ``out-of-time'' electron hits in the photon tagger~\cite{tagger-note}.

Outgoing particles were measured with the CLAS detector~\cite{CLAS}. This was a large-acceptance spectrometer, based on a toroidal magnet made of six superconducting coils arranged symmetrically around the beamline~\cite{CLAS-magnet}. The momentum of a charged particle was determined from the radius of curvature of its trajectory in the magnetic field as measured by a multi-wire drift-chamber system (DC)~\cite{DC}. A set of plastic scintillator counters (TOF), installed behind the drift chambers, provided the time of flight of each particle~\cite{TOF}. Particle identification was performed through the $\beta$ vs. $p$ technique. The energies and angles of the photons were measured with a lead/scintillator electromagnetic calorimeter (EC), covering polar angles in the range $8^\circ$--$45^\circ$, with energy resolution $\sigma_E/E\approx 10\%/\sqrt{E(\gev)}$, and angular resolution $\sigma_\theta \approx 10 \text{ mrad}$~\cite{EC}. 

The incoming photon was identified based on a $\pm~1.0$ ns coincidence between the vertex times obtained from the photon tagger and from the CLAS detector. The latter was determined by measuring the time of the outgoing charged particles with an array of plastic scintillator counters (ST) surrounding the target~\cite{ST}. Due to the large photon flux, a fraction $f_{multi-\gamma}=12.5\%$ of events with more than one tagged photon within the coincidence window was observed. To avoid any bias in the analysis, these events were discarded. This effect was accounted for in the cross-section normalization by scaling the measured event yield by $1/(1-f_{multi-\gamma})$. The systematic uncertainty of this correction, evaluated from the run-by-run variation of $f_{multi-\gamma}$, is $\approx 0.7\%$.
%including accidental photon tagger hits in the Monte Carlo simulations used to determine the detector acceptance and adopting the same selection procedure.

The trigger condition required one charged particle and two photons in the CLAS detector. The corresponding efficiency was evaluated from minimum bias runs and found to be on average $\varepsilon_\text{trg}=80\%$. A trigger efficiency map was derived and used to correct the cross-section normalization for the residual efficiency dependence on the charged particle impact point on the detector.

This analysis focuses on the $\gamma p \rightarrow \pi^0\eta p$ reaction, with all three final-state hadrons measured. Although CLAS was optimized for charged multi-particle final states, this reaction could be measured thanks to the high statistics and the specific setup of the \texttt{g12} run, with the target moved upstream to maximize the detector acceptance. Events were selected requiring detection of both the proton and the four photons from the $\pi^0$ and $\eta$ decay. 
The standard \gtw~procedures, including momentum corrections and fiducial cuts, were applied~\cite{g12-note}. A $4C$ kinematic fit (energy and momentum conservation imposed) was used to select events belonging to the exclusive $\gamma p \rightarrow 4\gamma p$ reaction, by introducing a cut on the corresponding confidence level (CL)~\cite{kfnote,kfnote2}. To optimize this cut, the difference between the missing mass on the proton squared and the four photon invariant mass squared -- here denoted as $K$ -- was considered.
%the $K$ kinematic variable, defined as the difference between the missing mass on the proton squared and the four photon invariant mass squared:
%\begin{equation}
%    K=\left(p^\mu_\text{beam}+p^\mu_\text{target}-p^\mu_p\right)^2-\left(\sum_{i=1}^4p^\mu_{\gamma_i}\right)^2 \,, 
%\end{equation}
%was introduced, where $p^\mu_j$ is the measured four momentum of particle $j$.
From energy and momentum conservation, it follows that signal events ($\gamma p \rightarrow 4\gamma p$) are distributed around $K=0$ with a gaussian distribution, while background events ($\gamma p \rightarrow 4\gamma pX$) manifest as a tail in the $K>0$ region. Therefore, the following figure of merit (FOM) was defined:
\begin{equation}
    \text{FOM} =\frac{n_s}{\sqrt{n_s+n_b}} \, \, ,
\end{equation}
where $n_s/2$ ($n_s/2+n_b$) was the number of events with $K<0$ ($K>0$). 
The optimal CL cut was determined by maximizing the FOM, and found to be $1.86\%$.
\begin{figure}
    \centering
    \includegraphics[width=.5\textwidth]{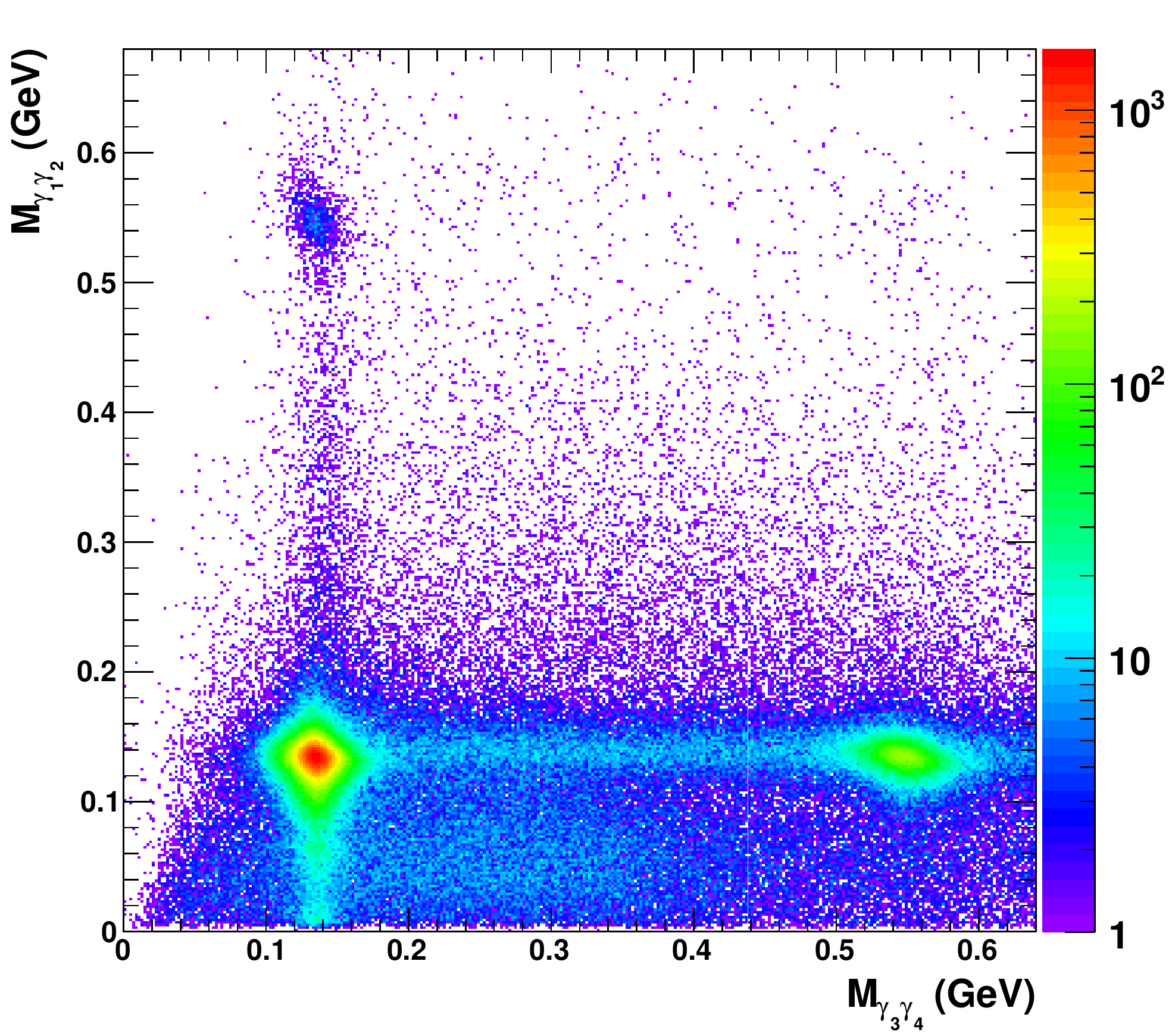}
    \caption{Correlation between the invariant mass of the two photon pairs for exclusive $\gamma p \rightarrow 4\gamma \,p$ events. 
    In each event, $\gamma_1$ and $\gamma_2$ are the photons with the smallest opening angle.
    The bottom-right cluster contains signal events from the $\gamma p \rightarrow \pi^0 \eta p$ reaction.
    %The few events appearing in the opposite photon assignment combination (top-left cluster) were not used in this analysis.
    }
    \label{fig:4gamma}
\end{figure}

%The kinematic fit discussed above was used to select a clean sample of $\gamma p \rightarrow 4\gamma p$ exclusive events.
The following procedure was then adopted to isolate the $\gamma p\rightarrow \pi^0 \eta p$ reaction.
First, the photons were ordered event-by-event by naming $\gamma_1$ and $\gamma_2$ those with the smallest opening angle. This algorithm exploits the fact that, due to the lower $\pi^0$ mass, the two photons from its decay are expected to have, on average, a smaller opening angle than those from $\eta$ decay. The corresponding efficiency, estimated from Monte Carlo simulations, is approximately $82\%$~\cite{Celentano_thesis}.
The correlation between the invariant masses of the two photon pairs, $M_{\gamma_1 \gamma_2}$ vs. $M_{\gamma_3 \gamma_4}$ is shown in  Fig.~\ref{fig:4gamma}. Signal events were identified as those corresponding to the bottom-right cluster centered at  $M_{\gamma_1 \gamma_2}=M_{\pi^0}, M_{\gamma_3 \gamma_4} = M_\eta$. A small fraction of events, corresponding to $\approx4\%$ of the main signal yield, appeared in the opposite combination, and was not considered in the following. 

After ordering the photons, the $M_{\gamma_3 \gamma_4}$ distribution showed a clear peak corresponding to the $\eta$, with some residual background events underneath.
%coming from exclusive $\gamma p \rightarrow 4\gamma p$ events with a different topology.
To reject these and extract the signal yield, the \sPlot method was used~\cite{splot}. This considers that events in the data sample originate from different independent sources and are characterized by a set of kinematic variables that can be split into two components. The method allows to reconstruct for each event source the distributions of \textit{control variables} from the knowledge of the Probability Density Funtion (PDF) associated to independent \textit{discriminating variables}. In this analysis, the invariant mass $M_{\gamma_3 \gamma_4}$ was used discriminating variable, while $M$ and $M_{\gamma_1 \gamma_2}$ were used as control variables. Two event sources were assumed: a signal source corresponding to the $\eta$ meson decay, modeled with a Gaussian PDF with exponential tails, and a background source, parameterized with a polynomial PDF. %The nominal fit range was $0.4\gev<M_{\gamma_3\gamma_4}<0.7\gev$. Events out of this range were rejected.
To avoid any correlation between variables that was induced by the kinematic fit, resulting in a possible bias, events were first divided into independent $M$ bins, and the \sPlot analysis was applied independently in each of them. To assess the quality of the result, the $M_{\gamma_1 \gamma_2}$ distribution for the signal source was investigated, finding that no residual background was present below the $\pi^0$ peak.

The CLAS acceptance and efficiency were evaluated by means of Monte Carlo simulations, based on a GEANT code that included knowledge of the full detector
geometry and a realistic response to traversing particles. 
Since the extracted differential cross section is integrated over some of the independent kinematic variables, such as the $\pi^0$ angles in the Gottfried-Jackson frame ($\Omega_{GJ}$), the model used to generate Monte Carlo events had to be as close as possible to the real physical one. To this end, $\gamma p \rightarrow \pi^0\eta p$ events were first generated according to a bremsstrahlung photon beam energy spectrum, with a phase-space distribution, and reconstructed through the same procedure used for real data. The result was used to compute the acceptance-corrected event distribution, from which a new Monte Carlo sample was generated. The procedure was iterated until a good agreement between data and  Monte Carlo was found for $-t$ and for $\Omega_{GJ}$ in each $E_{beam}$ bin. In particular, the good matching between data and Monte Carlo for $\Omega_{GJ}$ ensures that interference effects between different amplitudes contributing to the $\pi^0\eta$ final state is properly considered when computing the detector acceptance.
Finally, to account for the effect of the analysis procedures in the cross-section normalization, the same methods were applied to Monte Carlo events.
%These include the exclusion of events with more than one photon in the coincidence time window, the kinematic fit CL cut, the photon ordering algorithm, and the \sPlot method.

\begin{figure*}[t!]
    \centering
    \includegraphics[width=\textwidth]{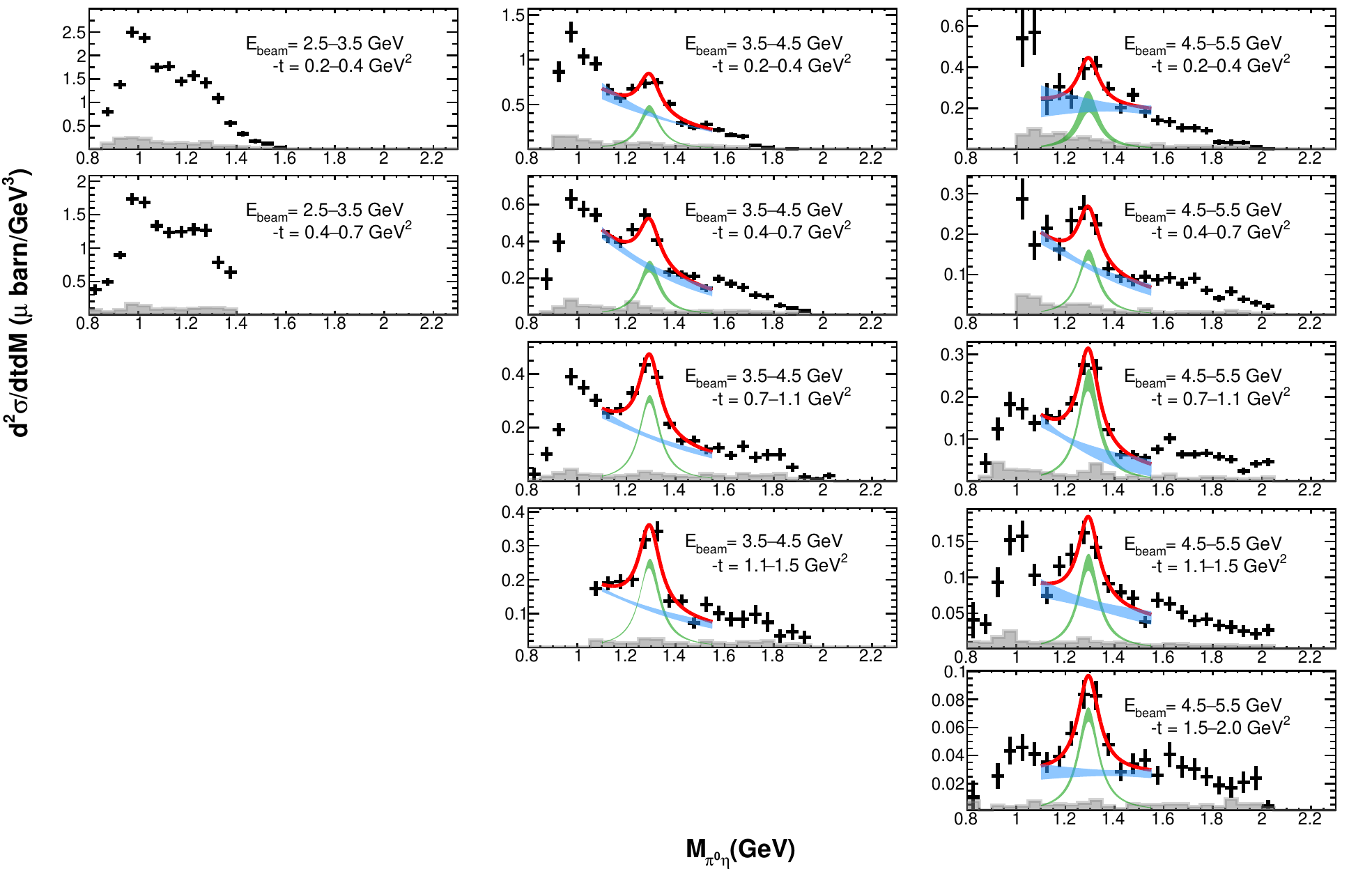}
    \caption{Differential cross section for the reaction $\gamma p \rightarrow \pi^0 \eta p$. Each histogram reports the reaction differential cross section $d^2\sigma/dtdM$  as a function of the $\pi^0 \, \eta$ invariant mass, for the specific $E_{beam}$ and $-t$ bin reported in the same panel.
    The bottom gray-filled area in each panel shows the systematic uncertainty. The red curve is the result from the best fit performed with the model described in the text. The green and blue areas correspond, respectively, to the contribution of the $a_2$ resonance and of the background,
    here reported as the $\pm 1\sigma$ systematic uncertainty bands around the central value.
    \textit{These have been scaled vertically by a factor $\times 2$ for better readability.}}
    \label{fig:xs}
\end{figure*}

The differential cross section $d^2\sigma/dt dM$ is shown in Fig.~\ref{fig:xs}, as a function of $M$, for three photon-beam energy bins (rows) and five four-momentum transfer bins (columns). The error bars report the statistical uncertainty only. %This was computed, in each bin, by adding in quadrature the statistical uncertainty on the event yield with that on the detector acceptance and efficiency.
Table~\ref{tab:systematics} summarizes the systematic uncertainties. The first four contributions are connected, respectively, to the uncertainty in the LH$_2$ target properties (density and length), the absolute photon flux normalization, the trigger system efficiency, and the $\eta \rightarrow \gamma \gamma$ branching fraction. 
The systematic uncertainties associated with the kinematic fit and the \sPlot procedure have been evaluated by
considering, in each bin, the relative variation of the cross section for different choices of the CL cut and of the degree of the background polynomial PDF. Finally, the systematic uncertainty on the CLAS acceptance was evaluated by varying the distribution used to generate the Monte Carlo events: a phase-space distribution was used, leading to a conservative estimate of this uncertainty contribution.
The total systematic uncertainty was obtained by adding in quadrature all individual terms.

\begin{table}[!t]
    \centering
    \begin{tabular}{|c|c|}
    \hline
    \textbf{Systematic Uncertainty Source} & \textbf{Magnitude} \\
    \hline
     Target properties    &  0.5$\%$ \\
     Photon flux & 5.7$\%$ \\
     Beam photon selection & 0.9$\%$ \\
     Trigger efficiency & 2.8$\%$ \\
     $\eta \rightarrow \gamma \gamma$ branching fraction & 0.5$\%$\\
     Kinematic fit & Variable, $\approx 3\%$ \\
     \sPlot & Variable,  $\approx 4\%$ \\
    Acceptance correction & Variable,  $\approx 5\%$ \\
    \hline
    \end{tabular}
    \caption{Summary of the systematic effects associated with the $\gamma p \rightarrow p\pi^0 \eta$ differential cross-section measurement. The effects marked as ``variable'' have a different contribution for each $E_\text{beam}$, $t$ and $M$ kinematic bin. The typical values are reported.}
    \label{tab:systematics}
\end{table}

The differential cross section $d^2\sigma / dtdM$ shows two distinctive structures corresponding to the $a_0(980)$ and $a_2(1320)$ resonances. In particular, the $a_2$ meson is clearly visible as a peak over a smooth background, with the latter decreasing at larger beam energies. The exclusive $a_2(1320)$ photoproduction cross section $d\sigma/dt$ has been extracted in the two largest photon beam energy bins by modeling $d^2\sigma/dtdM$ in the $M$ range $1.1$--$1.55$~$\gev$ as the incoherent sum of a  resonance term and a smooth background, including contributions from both non-resonant $\pi^0\eta$ photoproduction and from the residual high-mass tail of the $a_0(980)$ state.
The resonance term was written as the product of a $(E_\text{beam},-t)$--dependent production coefficient and a Breit-Wigner function that describes the $a_2$ line shape~\cite{PDG}.
The background term was parameterized as decreasing exponential function.
The cross-section model was convoluted with the experimental $M$ resolution, evaluated from Monte Carlo simulations. This ranged from a few \mev at high $M$ values up to $\approx 20$ $\mev$ at $M\approx 0.8$ $\gev$. A simultaneous $\chi^2$ fit to all $d^2\sigma/dtdM$ data points was then performed, with a total of 28 free parameters (9 $a_2$ production coefficients, 9 background polynomial terms, 9 background exponential slopes, and the $a_2$ mass).
In the Breit-Wigner formula, the $a_2$ mass $M_{a_2}$ was left to vary as a free parameter while the width $\Gamma_{a_2}$ was fixed to the nominal PDG value, $(113.4\pm1.3$) $\mev$ -- the effect of this choice was studied and included in the systematic uncertainty.
The $\chi^2/\text{NDF}$ value was $64.3/53=1.21$, and the obtained $M_{a_2}$ value was $(1308\pm2)$ $\mev$, in very good agreement with the nominal PDG value, $(1312.2\pm2.8)$ $\mev$. The fit result is reported for each kinematic bin in Fig.~\ref{fig:xs} as a red curve, while the green (blue) areas shows the $a_2$ (background) contribution only, reported as the $\pm 1 \sigma$ systematic uncertainty band around the central value.

\begin{figure}[!t]
    \centering
    \includegraphics[width=.45\textwidth]{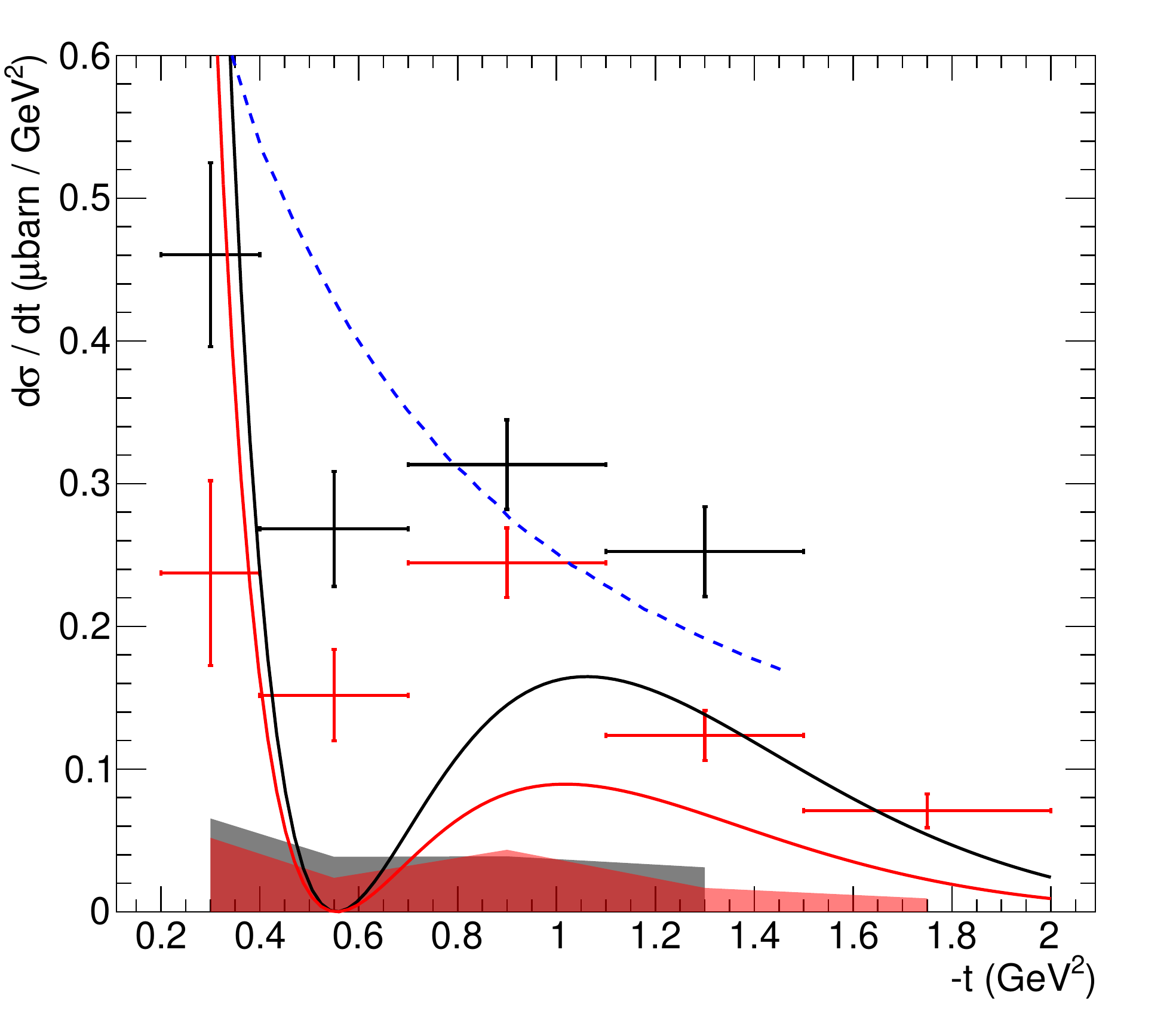}
    \caption{Differential cross section $d\sigma/dt$ for the reaction $\gamma p \rightarrow a_2(1320)p$, for $E_\text{beam}=3.5$--$4.5\gev$ (black) and $E_\text{beam}=4.5$--$5.5\gev$ (red). The vertical error bars show the statistical uncertainty, whereas horizontal error bars correspond to the $-t$ bins width. The bottom bands show the systematic uncertainty. The continuous lines are predictions from the JPAC model~\cite{Mathieu:2020zpm}, computed respectively for a beam energy of $4\gev$ (black) and $5\gev$ (red). The blue dashed line is the prediction from the model by Xie \textit{et al.}~\cite{Xie:2015isa}, for beam energy 3.4 GeV. \textit{For better readability, this was scaled vertically by a factor $\times 0.5$.}}
    \label{fig:a2}
\end{figure}

The differential cross section for the reaction $\gamma p \rightarrow a_2(1320)^0\,p$  was finally obtained by integrating the resonance term in each kinematic bin, accounting for the $a_2\rightarrow \pi^0 \eta$ branching fraction, $(14.5\pm1.2)\%$~\cite{PDG}. The results are shown in Fig.~\ref{fig:a2}, where the black (red) points refer to the photon energy range $3.5$--$4.5\gev$ ($4.5$--$5.5\gev$). For each data point, the vertical bar shows the statistical uncertainty, evaluated from the covariance matrix of the $\chi^2$ fit. The colored bands at the bottom show the systematic uncertainty, obtained summing quadratically the systematic uncertainty for $d^2\sigma/dtdM$ and that associated with the fit procedure. This was evaluated by repeating the fit with different choices of the fit range and of the $a_2$  width, that was varied within $\pm 2\sigma$ around the nominal value. $M_{min}$ ($M_{max}$) was varied in the interval $1.0$--$1.1$ GeV ($1.55$--$1.7$ GeV). The nominal range reported previously corresponds to the fit with the smallest $\chi^2$/NDF value. The
argument of the exponential function was also replaced by polynomials of
various orders. The systematic uncertainty was calculated, in each bin, as the RMS of the cross-section values obtained from the different fits.

The most intriguing feature of the $\gamma p \rightarrow  a_2(1320)\,p$ cross section is the  presence of a dip at $-t_{dip}\approx0.55\gevsq$, observed simultaneously at both beam energies. The hypothesis that this observed was just the effect of a statistical fluctuation was excluded at 99$\%$ CL as follows. 
We made a null hypothesis for $d\sigma/dt$, assuming a monotone shape: we tested both a linear and an exponential behavior. In particular, for this $-t$ and $E_{beam}$ range, within the precision dictated by the large statistical errors, the latter functional form should generally
provide a sufficient description of 
$\frac{d\sigma}{dt}$ in the absence of a dip. We generated $N=10^5$ toy Monte Carlo datasets, re-sampling each measured $d\sigma/dt$ point $y_i \pm \sigma_i$ from a Gaussian distribution with $\mu=y_i$ and $\sigma=\sigma_i$. Conservatively, we adopted for $\sigma_i$ the quadratic sum of the statistical and systematic uncertainties, excluding the contributions that are indepedentent from the kinematic bin. For each toy dataset and each beam energy, we performed a fit with the hypothesized functional form, excluding the point in the dip, extrapolating from it the expected cross-section value at $-t_{dip}$. The bin width was taken into account by computing the latter as the average cross-section value inside the $-t_{dip}$ bin.
Finally, from the fraction of toy datasets in which both extrapolated values were lower than the toy dataset values at $-t_{dip}$, we computed the null hypothesis $p-$value.

The origin of the dip and its specific location can be explained in the context of Regge theory~\cite{Irving:1977ea}. In Fig.~\ref{fig:a2}, we show the results of a model based on a Regge-theory production amplitude parametrization developed by the JPAC Collaboration~\cite{Mathieu:2020zpm}, computed for the two beam energies 4 GeV (black) and 5 GeV (red). 
The amplitude includes the leading vector trajectories only, which have the $\rho$ and $\omega$ quantum numbers. 
%, and its magnitude is determined by the  coupling  of the vector trajectory to $\gamma a_2$. This is computed using Vector Meson Dominance, from the known $a_2 \to \omega\pi\pi$ width~\cite{PDG}, by further assuming that the $\rho$ dominates the $\pi\pi$ state. 
Regge-resonance duality implies that the parameters of Regge amplitudes corresponding to these vector 
  exchanges  are closely related to the ones involving the tensor $a_2$ and $f_2$ mesons (exchange degeneracy hypothesis~\cite{Mandula:1970wz, Irving:1977ea}).
 Since no scalar mesons lie on the $a_2$ trajectory, the residue of the tensor exchange has to vanish when the Regge trajectory $\alpha(t)$ is equal to zero to remove the scalar pole. Vector exchanges, which share the residues with the tensors,  will thus also vanish at $\alpha(t)=0$, that is at $-t= m^2_{\rho,\omega} \approx 0.55\gevsq$, leading to 
  an exact zero in the cross section. However, subleading Regge poles or cut contributions can turn the zero of the amplitude into the dip observed in data and improve the description at higher $-t$. The results presented here are a pure prediction for $d\sigma/dt$, since the model parameters were tuned on different datasets: the qualitative agreement between data and model, in particular concerning the position of the dip, demonstrates the effectiveness of a reaction amplitude parametrization based on Regge phenomenology. The use of the present data to fine-tune the model parameters is beyond the scope of this work, and will be the subject of a different publication~\cite{Mathieu:2020zpm}. Finally, we observe that our new data will help in understanding the nature of the $a_2(1320)$ resonance.  While many authors describe it as a $q\overline{q}$ state~\cite{Cirigliano:2003yq}, others propose a different description. For example, Xie \textit{et al.}~\cite{Xie:2015isa} recently developed a model where the $a_2$ is a molecular state dynamically generated from the $\rho-\omega$ and $\rho-\phi$ interactions in $S$-wave with spin 2. This model predicts a smooth $d\sigma/dt$ shape, without any dip. Our data rules out this hypothesis. 

%Summary
In summary, we have measured for the first time the reaction $\gamma p \rightarrow \pi^0 \eta p$ in the photon beam energy range $3.5$--$5.5\gev$, and for four-momentum transferred squared values between $0.2$ and $2.0\gevsq$, extracting the cross section for the exclusive $a_2(1320)$ photoproduction on the proton. The cross section shows a pronounced dip at $-t\approx 0.55 \gevsq$, which can be explained in the framework of Regge theory. 
Since the $a_2(1320)^0$ is the most prominent structure present in the $\pi^0\,\eta$ invariant mass, detailed knowledge of its production cross section is valuable for any assessment of a possible exotic resonance contribution. This measurement will thus help high statistics photoproduction experiments, \textit{e.g.} CLAS12~\cite{Celentano_2013}, GLUEX~\cite{Austregesilo:2018mno}, and BGOOD~\cite{Alef:2019imq}, to better understand the $\pi^0\,\eta$ mass spectrum and to properly describe the production of the dominant $a_2$ resonance, using it as a benchmark in the search for exotic states.

\begin{acknowledgments}
This work was supported by: the U.S. Department of Energy (DOE), the U.S. National Science Foundation, the U.S. Jeffress Memorial Trust; the Physics and Astronomy Department and the Office of Research and Economic Development at Mississippi State University, the United Kingdom's Science and Technology Facilities Council (STFC), the Italian Istituto Nazionale di Fisica Nucleare; the French Institut National de Physique Nucl\'eaire et de Physique des Particules, the French Centre National de la Recherche Scientifique; and the National Research Foundation of Korea. This material is based upon work supported by the U.S. Department of Energy, Office of Science, Office of Nuclear Physics under contract DE-AC05-06OR23177.
V.M. is supported by Comunidad
Aut\'onoma de Madrid through Programa de Atracci\'on de Talento Investigador 2018 (Modalidad 1). 
\end{acknowledgments}
%\bibliographystyle{apsrev4-2}
%\bibliography{biblio}% Produces the bibliography via BibTeX.

\providecommand{\noopsort}[1]{}\providecommand{\singleletter}[1]{#1}%

\end{document}

%% file: authors.tex
\newcommand*{\ANL}{Argonne National Laboratory, Argonne, Illinois 60439}
\newcommand*{\ANLindex}{1}
%\affiliation{\ANL}
\newcommand*{\CSUDH}{California State University, Dominguez Hills, Carson, CA 90747}
\newcommand*{\CSUDHindex}{2}
%\affiliation{\CSUDH}
\newcommand*{\CANISIUS}{Canisius College, Buffalo, NY}
\newcommand*{\CANISIUSindex}{3}
%\affiliation{\CANISIUS}
\newcommand*{\CMU}{Carnegie Mellon University, Pittsburgh, Pennsylvania 15213}
\newcommand*{\CMUindex}{4}
%\affiliation{\CMU}
\newcommand*{\CUA}{Catholic University of America, Washington, D.C. 20064}
\newcommand*{\CUAindex}{5}
%\affiliation{\CUA}
\newcommand*{\SACLAY}{IRFU, CEA, Universit\'{e} Paris-Saclay, F-91191 Gif-sur-Yvette, France}
\newcommand*{\SACLAYindex}{6}
%\affiliation{\SACLAY}
\newcommand*{\CNU}{Christopher Newport University, Newport News, Virginia 23606}
\newcommand*{\CNUindex}{7}
%\affiliation{\CNU}
\newcommand*{\UCONN}{University of Connecticut, Storrs, Connecticut 06269}
\newcommand*{\UCONNindex}{8}
%\affiliation{\UCONN}
\newcommand*{\DUKE}{Duke University, Durham, North Carolina 27708-0305}
\newcommand*{\DUKEindex}{9}
%\affiliation{\DUKE}
\newcommand*{\DUQUESNE}{Duquesne University, 600 Forbes Avenue, Pittsburgh, PA 15282 }
\newcommand*{\DUQUESNEindex}{10}
%\affiliation{\DUQUESNE}
\newcommand*{\FU}{Fairfield University, Fairfield CT 06824}
\newcommand*{\FUindex}{11}
%\affiliation{\FU}
\newcommand*{\FERRARAU}{Universita' di Ferrara, 44121 Ferrara, Italy}
\newcommand*{\FERRARAUindex}{12}
%\affiliation{\FERRARAU}
\newcommand*{\FIU}{Florida International University, Miami, Florida 33199}
\newcommand*{\FIUindex}{13}
%\affiliation{\FIU}
\newcommand*{\FSU}{Florida State University, Tallahassee, Florida 32306}
\newcommand*{\FSUindex}{14}
%\affiliation{\FSU}
\newcommand*{\GWUI}{The George Washington University, Washington, DC 20052}
\newcommand*{\GWUIindex}{15}
%\affiliation{\GWUI}
\newcommand*{\ISU}{Idaho State University, Pocatello, Idaho 83209}
\newcommand*{\ISUindex}{16}
%\affiliation{\ISU}
\newcommand*{\INFNFE}{INFN, Sezione di Ferrara, 44100 Ferrara, Italy}
\newcommand*{\INFNFEindex}{17}
%\affiliation{\INFNFE}
\newcommand*{\INFNFR}{INFN, Laboratori Nazionali di Frascati, 00044 Frascati, Italy}
\newcommand*{\INFNFRindex}{18}
%\affiliation{\INFNFR}
\newcommand*{\INFNGE}{INFN, Sezione di Genova, 16146 Genova, Italy}
\newcommand*{\INFNGEindex}{19}
%\affiliation{\INFNGE}
\newcommand*{\INFNRO}{INFN, Sezione di Roma Tor Vergata, 00133 Rome, Italy}
\newcommand*{\INFNROindex}{20}
%\affiliation{\INFNRO}
\newcommand*{\INFNTUR}{INFN, Sezione di Torino, 10125 Torino, Italy}
\newcommand*{\INFNTURindex}{21}
%\affiliation{\INFNTUR}
\newcommand*{\INFNPAV}{INFN, Sezione di Pavia, 27100 Pavia, Italy}
\newcommand*{\INFNPAVindex}{22}
%\affiliation{\INFNPAV}
\newcommand*{\ORSAY}{Universit'{e} Paris-Saclay, CNRS/IN2P3, IJCLab, 91405 Orsay, France}
\newcommand*{\ORSAYindex}{23}
%\affiliation{\ORSAY}
\newcommand*{\Juelich}{Institute fur Kernphysik (Juelich), Juelich, Germany}
\newcommand*{\Juelichindex}{24}
%\affiliation{\Juelich}
\newcommand*{\JMU}{James Madison University, Harrisonburg, Virginia 22807}
\newcommand*{\JMUindex}{25}
%\affiliation{\JMU}
\newcommand*{\KNU}{Kyungpook National University, Daegu 41566, Republic of Korea}
\newcommand*{\KNUindex}{26}
%\affiliation{\KNU}
\newcommand*{\LAMAR}{Lamar University, 4400 MLK Blvd, PO Box 10009, Beaumont, Texas 77710}
\newcommand*{\LAMARindex}{27}
%\affiliation{\LAMAR}
\newcommand*{\MIT}{Massachusetts Institute of Technology, Cambridge, Massachusetts  02139-4307}
\newcommand*{\MITindex}{28}
%\affiliation{\MIT}
\newcommand*{\MISS}{Mississippi State University, Mississippi State, MS 39762-5167}
\newcommand*{\MISSindex}{29}
%\affiliation{\MISS}
\newcommand*{\ITEP}{National Research Centre Kurchatov Institute - ITEP, Moscow, 117259, Russia}
\newcommand*{\ITEPindex}{30}
%\affiliation{\ITEP}
\newcommand*{\UNH}{University of New Hampshire, Durham, New Hampshire 03824-3568}
\newcommand*{\UNHindex}{31}
%\affiliation{\UNH}
\newcommand*{\NSU}{Norfolk State University, Norfolk, Virginia 23504}
\newcommand*{\NSUindex}{32}
%\affiliation{\NSU}
\newcommand*{\OHIOU}{Ohio University, Athens, Ohio  45701}
\newcommand*{\OHIOUindex}{33}
%\affiliation{\OHIOU}
\newcommand*{\ODU}{Old Dominion University, Norfolk, Virginia 23529}
\newcommand*{\ODUindex}{34}
%\affiliation{\ODU}
\newcommand*{\ROMAII}{Universita' di Roma Tor Vergata, 00133 Rome Italy}
\newcommand*{\ROMAIIindex}{35}
%\affiliation{\ROMAII}
\newcommand*{\MSU}{Skobeltsyn Institute of Nuclear Physics, Lomonosov Moscow State University, 119234 Moscow, Russia}
\newcommand*{\MSUindex}{36}
%\affiliation{\MSU}
\newcommand*{\SCAROLINA}{University of South Carolina, Columbia, South Carolina 29208}
\newcommand*{\SCAROLINAindex}{37}
%\affiliation{\SCAROLINA}
\newcommand*{\TEMPLE}{Temple University,  Philadelphia, PA 19122 }
\newcommand*{\TEMPLEindex}{38}
%\affiliation{\TEMPLE}
\newcommand*{\JLAB}{Thomas Jefferson National Accelerator Facility, Newport News, Virginia 23606}
\newcommand*{\JLABindex}{39}
%\affiliation{\JLAB}
\newcommand*{\UTFSM}{Universidad T\'{e}cnica Federico Santa Mar\'{i}a, Casilla 110-V Valpara\'{i}so, Chile}
\newcommand*{\UTFSMindex}{40}
%\affiliation{\UTFSM}
\newcommand*{\INSUBRIA}{Universit\`{a} degli Studi dell'Insubria, 22100 Como, Italy}
\newcommand*{\INSUBRIAindex}{41}
%\affiliation{\INSUBRIA}
\newcommand*{\BRESCIA}{Universit\`{a} degli Studi di Brescia, 25123 Brescia, Italy}
\newcommand*{\BRESCIAindex}{42}
%\affiliation{\BRESCIA}
\newcommand*{\GLASGOW}{University of Glasgow, Glasgow G12 8QQ, United Kingdom}
\newcommand*{\GLASGOWindex}{43}
%\affiliation{\GLASGOW}
\newcommand*{\YORK}{University of York, York YO10 5DD, United Kingdom}
\newcommand*{\YORKindex}{44}
%\affiliation{\YORK}
\newcommand*{\VT}{Virginia Tech, Blacksburg, Virginia   24061-0435}
\newcommand*{\VTindex}{45}
%\affiliation{\VT}
\newcommand*{\VIRGINIA}{University of Virginia, Charlottesville, Virginia 22901}
\newcommand*{\VIRGINIAindex}{46}
%\affiliation{\VIRGINIA}
\newcommand*{\WM}{College of William and Mary, Williamsburg, Virginia 23187-8795}
\newcommand*{\WMindex}{47}
%\affiliation{\WM}
\newcommand*{\YEREVAN}{Yerevan Physics Institute, 375036 Yerevan, Armenia}
\newcommand*{\YEREVANindex}{48}
%\affiliation{\YEREVAN}

\newcommand*{\NOWOHIOU}{Ohio University, Athens, Ohio  45701}
\newcommand*{\NOWZZZ}{unused, unused}
\newcommand*{\NOWISU}{Idaho State University, Pocatello, Idaho 83209}
\newcommand*{\NOWBRESCIA}{Universit\`{a} degli Studi di Brescia, 25123 Brescia, Italy}
\newcommand*{\NOWJLAB}{Thomas Jefferson National Accelerator Facility, Newport News, Virginia 23606}

\newcommand*{\IPARCOS}{Departamento de Física Teórica and IPARCOS, Universidad Complutense de Madrid, 28040 Madrid, Spain}
 \newcommand*{\IPARCOSindex}{43}
 
 \newcommand*{\INDIANA}{Physics Department, Indiana University, Bloomington, IN 47405, USA}
 \newcommand*{\INDIANAindex}{44}
 
 \newcommand*{\CEEM}{Center for Exploration of Energy and Matter, Indiana University, Bloomington, IN 47403, USA}
 \newcommand*{\CEEMindex}{44}
 
 \newcommand*{\JLABTHEORY}{Theory Center, Thomas Jefferson National Accelerator Facility, Newport News, VA 23606, USA}
\newcommand*{\JLABTHEORYindex}{45}

\newcommand*{\ECT}{European Centre for Theoretical Studies and Nuclear Physics (ECT$^{*}$) and Fondazione Bruno Kessler,
Strada delle Tabarelle 286, Villazzano (Trento), I-38123, Italy}
\newcommand*{\ECTindex}{46}
\newcommand*{\ASU}{Arizona State University, Tempe, Arizona 85287-1504}
\newcommand*{\ASUindex}{47}

\newcommand*{\NOWINFNGE}{INFN, Sezione di Genova, 16146 Genova, Italy}

 %%%%%%%%%%%%%%% END OF Latex Macros for institute addresses  %%%%%%%%%%%%%%%%%%%%%%%%% 

\author {A.~Celentano} 
\affiliation{\INFNGE}
\author {M.~Battaglieri} 
\affiliation{\JLAB}
\affiliation{\INFNGE}
\author {R.~De~Vita} 
\affiliation{\INFNGE}
\author {L.~Marsicano} 
\affiliation{\INFNGE}
\author{V.~Mathieu}
\affiliation{\IPARCOS}
\author{A.~Pilloni}
\affiliation{\ECT}
\affiliation{\INFNGE}
\author{A.~Szczepaniak}
\affiliation{\INDIANA}
\affiliation{\CEEM}
\affiliation{\JLABTHEORY}
\author{K.~P.~Adhikari}
\affiliation{\ODU}
\author {S. Adhikari} 
\affiliation{\FIU}
\author {M.J.~Amaryan} 
\affiliation{\ODU}
% ...

\collaboration{The CLAS Collaboration}
\noaffiliation